\documentstyle[12pt]{article}
\newcommand{\bq}{\begin{equation}}
\newcommand{\eq}{\end{equation}}
\newcommand{\ba}{\begin{eqnarray}}
\newcommand{\ea}{\end{eqnarray}}

\newcommand{\nl }{ \nonumber  }

\newcommand{\ul}{\underline}
\newcommand{\p}{\partial}

\newcommand{\h}{\hspace{1cm}}
\newcommand{\s}{\sigma}
\newcommand{\us}{\underline\sigma}
\newcommand{\da}{\delta^p(\us_1 - \us_2)}
\newcommand{\la}{\lambda}
\newcommand{\La}{\Lambda}
\begin{document}
\begin{titlepage}
\vspace*{2cm}
{\bf\begin{center}
 N=1, D=10 TENSIONLESS SUPERBRANES I.
\footnote{Work supported in part by the National Science Foundation
of Bulgaria under contract $\phi-620/1996$}
\vspace*{2cm}
\\
P. Bozhilov
\footnote {permanent address:
\it Dept.of Theoretical Physics,"Konstantin Preslavski" Univ. of Shoumen
\\
9700 Shoumen, Bulgaria  E-mail: bojilov@uni-shoumen.bg}
\\
\it
Bogoliubov Laboratory of Theoretical Physics, JINR, Dubna, Russia
\\
\it
E-mail: bojilov@thsun1.jinr.dubna.su
\vspace*{5cm}
\end{center}}

We consider a model for tensionless (null) super $p$-branes in the
Hamiltonian approach and in the framework of a harmonic superspace.
The obtained algebra of Lorentz-covariant, irreducible, first class
constraints is such that the BRST charge corresponds to a first 
rank system.  
\end{titlepage} 
\normalsize 
\vspace*{2cm}
\section{\bf Introduction}

\hspace{1cm}
The tensionless (null) $p$-branes correspond to usual $p$-branes with their
tension
\ba\nl
T=(2\pi\alpha^\prime)^{-(p+1)/2}
\ea
taken to be zero. This relationship between null $p$-branes and the
tensionful ones may be regarded as a generalization of the massless-massive
particles correspondence. On the other hand, the limit $T\to 0$ 
(because of $(\alpha^\prime)^{-1}\propto M^2_{Plank}$) corresponds 
to the energetic scale $E>>M_{Plank}$. In other words, the null 
$p$-brane is the high energy limit of the tensionful one. There 
exist also an interpretation of the null and free $p$-branes as 
theories, corresponding to different vacuum states of a $p$-brane, 
interacting with a scalar field background \cite{BZ}. So, one can 
consider the possibility of tension generation for null $p$-branes
(see \cite{BSTV} and references therein). Another viewpoint on the connection
between null and tensionful $p$-branes is that the null one may be
interpreted as a "free" theory opposed to the tensionful "interacting"
theory \cite{G}. All this explains the interest in considering null
$p$-branes and their supersymmetric extensions.

   Models for tensionless $p$-branes with manifest supersymmetry
are proposed in \cite{Zh}. In \cite{BZ} a twistor-like action
is suggested, for null super-$p$-branes with $N$-extended global
supersymmetry in four dimensional space-time. Then, in the framework of the
Hamiltonian formalism, the initial algebra of first and second class
constraints is converted into an algebra of first class effective constraints
only. The obtained BRST charge corresponds to second rank theory. It is
proven that there are no quantum anomalies when the so called "generalized"
$\hat q\hat p$ operator ordering is applied.
In the recent work \cite{S}, among other problems, the quantum
constraint algebras of the usual and conformal tensionless spinning
$p$-branes are considered.

In a previous paper \cite{136}, we announced for a null super 
$p$-brane model which possesses the following classical constraint 
algebra 
\ba \nl 
\{T_0(\ul \sigma_1),T_0(\ul \sigma_2)\}&=&0, 
\\ \nl 
\{T_0(\ul \sigma_1),T_{\alpha}^{A}(\ul \sigma_2)\}&=&0 ,
\\
\nl
\{T_0(\ul \sigma_1),T_{j}^{A}(\ul \sigma_2)\}&=&
[T_0(\ul \sigma_1) + T_0(\ul \sigma_2)]
\p_j \delta^p (\ul \sigma_1 - \ul \sigma_2) ,
\\
\label{1}
\{T_{j}^{A}(\ul \sigma_1),T_{k}^{B}(\ul \sigma_2)\}&=&
\delta^{AB}[\delta_{j}^{l}T_{k}^{B}(\ul \sigma_1) +
\delta_{k}^{l}T_{j}^{A}(\ul \sigma_2)]\p_l\delta^p(\ul \sigma_1-\ul \sigma_2) ,
\\
\nl
\{T_{j}^{A}(\ul \sigma_1),T_{\alpha}^{B}(\ul \sigma_2)\}&=&
\delta^{AB}[T_{\alpha}^{A}(\ul \sigma_1)+T_{\alpha}^{A}(\ul \sigma_2)]
\p_j \delta^p (\ul \sigma_1-\ul \sigma_2) ,
\\
\nl
\{T_{\alpha}^{A}(\ul \sigma_1),T_{\beta}^{B}(\ul \sigma_2)\}&=&
-2i\delta^{AB}\hat{P}_{\alpha\beta}T_0(\ul \sigma_1)
\delta^p (\ul \sigma_1-\ul \sigma_2) ,
\\
\nl
\hat{P}_{\alpha\beta}&=&P_{\mu}\sigma_{\alpha\beta}^{\mu} .
\ea
Here $\ul\sigma=(\sigma_1, ... ,\sigma_p)$, $(j,k=1, ... , p)$,
$(A,B=1, ... ,N)$,
where N is the number of the supersymmetries, $\alpha, \beta$ are spinor
indices and $P_{\mu}$ is a Lorentz vector $(\mu=0,1,..,D-1)$. In (\ref{1})
and below we do not write explicitly the dependence of the 
quantities on the time parameter $\tau$.

In this letter, we consider the particular case of $N=1$, $D=10$ 
tensionless superbrane. We work in the Hamiltonian approach and in 
the framework of a harmonic superspace. Passing to a system of 
Lorentz-covariant, irredusible first class constraints, we obtain a 
BRST charge as for a first rank theory.  
\section{\bf Hamiltonian formulation}
\hspace{1cm}
Let us begin with writing the initial Hamiltonian of the dynamical system
under consideration
\ba \nl
H_0=\int d^p\sigma\bigl [\mu^0 T_0+\mu^j T_j+\mu^\alpha D_{\alpha}\bigr ]
\ea
$H_0$ is a linear combination of the constraints $T_0(\us)$, $T_j(\us)$ and
$D_{\alpha}(\us)$. The latter are given by the expressions:
\ba
\nl
T_0 &=& p_{\mu}p_{\nu}\eta^{\mu\nu},
\h
diag(\eta_{\mu\nu})=(-,+,...,+), \\
\label{con}
T_j &=& p_{\nu}\p_j x^{\nu}+p_{\theta\alpha}\p_j \theta^{\alpha},
\h
\p_j=\p/\p\s^j ,
\\
\nl
D_{\alpha}&=& -ip_{\theta\alpha}-(\not{p}\theta)_{\alpha}.
\ea
Here $(x^{\nu}$, $\theta^{\alpha})$ are the superspace coordinates,
$\theta^{\alpha}$ is a left Majorana-Weyl spinor $(\alpha=1,...,16)$,
$p_{\nu}$, $p_{\theta\alpha}$ are the corresponding conjugated momenta,
$\not p = p_{\mu}\sigma^{\mu}$, where $\sigma^{\mu}$ are the 10-dimensional
$\sigma$-matrices (for our conventions see the Appendix).

$H_0$ generalizes on the one hand the bosonic null $p$-brane Hamiltonian
\ba
\nl
H^B=\int d^p\s\bigl (\mu^0 p^2+\mu^j p_\nu\p_j x^\nu\bigr ),
\ea
and on the other - the $N=1$ Brink-Shwarz superparticle with Hamiltonian
\ba
\nl
H^{BS}=\mu^0 p^2+\mu^\alpha D_\alpha .
\ea

The constraints (\ref{con}) satisfy the following (equal $\tau$) Poisson
bracket algebra
\ba
\nl
\{T_0(\ul \sigma_1),T_0(\ul \sigma_2)\}&=&0,
\\
\nl
\{T_0(\ul \sigma_1),D_{\alpha}(\ul \sigma_2)\}&=&0 ,
\\
\nl
\{T_0(\ul \sigma_1),T_{j}(\ul \sigma_2)\}&=&
[T_0(\ul \sigma_1) + T_0(\ul \sigma_2)]
\p_j \delta^p (\ul \sigma_1 - \ul \sigma_2) ,
\\ \nl
\{T_{j}(\ul \sigma_1),T_{k}(\ul \sigma_2)\}&=&
[\delta_{j}^{l}T_{k}(\ul \sigma_1) +
\delta_{k}^{l}T_{j}(\ul \sigma_2)]\p_l\delta^p(\ul \sigma_1-\ul \sigma_2) ,
\\
\nl
\{T_{j}(\ul \sigma_1),D_{\alpha}(\ul \sigma_2)\}&=&
D_{\alpha}(\ul \sigma_1)
\p_j \delta^p (\ul \sigma_1-\ul \sigma_2) ,
\\
\nl
\{D_{\alpha}(\ul \sigma_1),D_{\beta}(\ul \sigma_2)\}&=&
2i\not{p}_{\alpha\beta}
\delta^p (\ul \sigma_1-\ul \sigma_2) .
\ea
From the condition that the constraints must be maintained in time,
i.e. \cite{D}
\ba\label{Dc}
\bigl \{T_0,H_0\bigr \}\approx 0,
\bigl \{T_j,H_0\bigr \}\approx 0,
\bigl \{D_\alpha,H_0\bigr \}\approx 0,
\ea
it follows that in the Hamiltonian $H_0$ one has to include the constraints
\ba
\nl
T_{\alpha}=\not p_{\alpha\beta} D^{\beta}
\ea
instead of $D_{\alpha}$. This is because the Hamiltonian has to be first
class quantity, but $D_{\alpha}$ are a mixture of first and second class
constraints. $T_{\alpha}$ has the following non-zero Poisson brackets
\ba
\nl
\{T_{j}(\ul \sigma_1),T_{\alpha}(\ul \sigma_2)\}&=&
[T_{\alpha}(\ul \sigma_1)+T_{\alpha}(\ul \sigma_2)]
\p_j \delta^p (\ul \sigma_1-\ul \sigma_2) ,
\\ \nl
\{T_{\alpha}(\ul \sigma_1),T_{\beta}(\ul \sigma_2)\}&=&
2i\not{p}_{\alpha\beta}T_0
\delta^p (\ul \sigma_1-\ul \sigma_2) .
\ea
In this form, our constraints are first class
(their algebra coincides with the algebra (\ref{1}) for
$N=1$ and $P_{\mu}=-p_{\mu}$) and the Dirac consistency conditions 
(\ref{Dc}) (with $D_\alpha$ replaced by $T_\alpha$) are satisfied 
identically.  However, one now encounters a new problem. The 
constraints $T_0$, $T_j$ and $T_{\alpha}$ are not irreducible, i.e. 
they are functionally dependent:  \ba \nl (\not p T)^{\alpha} - 
D^{\alpha} T_0 = 0 .  \ea It is known, that in this case after 
BRST-BFV quantization an infinite number of ghosts for ghosts 
appear, if one wants to preserve the manifest Lorentz invariance. 
The way out consists in the introduction of auxiliary variables, so 
that the mixture of first and second class constraints $D^{\alpha}$ 
can be appropriately covariantly decomposed into first class 
constraints and second class ones. To this end, here we will use 
the auxiliary harmonic variables introduced in \cite{Sok} and 
\cite{NissP}.  These are $u_{\mu}^a$ and $v_{\alpha}^{\pm}$, where 
superscripts $a=1,...,8$ and $\pm$ transform under the 'internal' 
groups $SO(8)$ and $SO(1,1)$ respectively. The just introduced 
variables are constrained by the following orthogonality conditions 
\ba\nl
u_{\mu}^a u^{b\mu}=C^{ab},
\h
u_{\mu}^{\pm}u^{a\mu}=0,
\h
u_{\mu}^{+} u^{-\mu}=-1,
\ea
where
\ba\nl
u_{\mu}^{\pm}=v_{\alpha}^\pm \sigma_{\mu}^{\alpha\beta}v_{\beta}^\pm ,
\ea
$C^{ab}$ is the invariant metric tensor in the relevant representation
space of $SO(8)$ and $(u^\pm)^2=0$ as a consequence of the Fierz identity
for the 10-dimensional $\sigma$-matrices. We note that $u^a_\nu$ and
$v_{\alpha}^\pm$ do not depend on $\us$.

Now we have to ensure that our dynamical system does not depend on
arbitrary rotations of the auxiliary variables $(u_{\mu}^a$, $u_{\mu}^\pm)$.
It can be done by introduction of first class constraints, which
generate these transformations
\ba\nl
I^{ab}&=&-(u_{\nu}^a p^{b\nu}_u - u_{\nu}^b p^{a\nu}_u +
\frac{1}{2}v^+\sigma^{ab}p_v^+ +\frac{1}{2}v^-\sigma^{ab}p_v^-),
\h
\sigma^{ab}=u_{\mu}^a u_{\nu}^b \sigma^{\mu\nu},
\\
\label{ncon}
I^{-+}&=&-\frac{1}{2}(v_{\alpha}^+ p_v^{+\alpha} -
v_{\alpha}^- p_v^{-\alpha}),
\\ \nl
I^{\pm a}&=&-(u_{\mu}^\pm p_u^{a\mu} +
\frac{1}{2}v^\mp \sigma^\pm \sigma^a p_v^\mp) ,
\h
\sigma^\pm =u^\pm_\nu \sigma^\nu,
\h
\sigma^a = u^a_\nu \sigma^\nu .
\ea
In the above equalities, $p_u^{a\nu}$ and $p_v^{\pm \alpha}$ are the momenta
canonically conjugated to $u^a_\nu$ and $v^\pm_\alpha$.

The newly introduced constraints (\ref{ncon}) obey the following Poisson
bracket algebra
\ba\nl
\{I^{ab},I^{cd}\}&=&C^{bc}I^{ad}-C^{ac}I^{bd}+C^{ad}I^{bc}-C^{bd}I^{ac},
\\ \nl
\{I^{-+},I^{\pm a}\}&=&\pm I^{\pm a},
\\ \nl
\{I^{ab},I^{\pm c}\}&=&C^{bc}I^{\pm a}-C^{ac}I^{\pm b},
\\ \nl
\{I^{+a},I^{-b}\}&=&C^{ab}I^{-+} + I^{ab} .
\ea
This algebra is isomorphic to the $SO(1,9)$ algebra: $I^{ab}$ generate $SO(8)$
rotations, $I^{-+}$ is the generator of the subgroup $SO(1,1)$ and
$I^{\pm a}$ generate the transformations from the coset
$SO(1,9)/SO(1,1)\times SO(8)$.

Now we are ready to separate $D^\alpha$ into first and second
class constraints in a Lorentz-covariant form. This separation is given
by the equalities \cite{NissPS}:
\ba\label{rev}
D^\alpha &=&\frac{1}{p^+}\bigl [(\sigma^a v^+)^\alpha D_a +
(\not p \sigma^+ \sigma^a v^-)^\alpha G_a \bigr ],
\h
p^+ = p^\nu u^+_\nu,
\\ \nl
D^a &=&(v^+ \sigma^a \not p )_\beta D^\beta,
\h
G^a = \frac{1}{2}(v^- \sigma^a \sigma^+)_\beta D^\beta.
\ea
Here $D^a$ are first class constraints and $G^a$ are second class ones:
\ba\nl
\{D^a(\us_1),D^b(\us_2)\}&=&-2iC^{ab}p^+ T_0 \delta^p(\us_1 - \us_2)
\\ \nl
\{G^a(\us_1),G^b(\us_2)\}&=&iC^{ab}p^+ \da .
\ea
It is convenient to pass from second class constraints $G^a$ to first
class constraints $\hat G^a$, without changing the actual degrees of freedom
\cite{NissPS}, \cite{EM} :
\ba\nl
G^a \rightarrow \hat G^a = G^a + (p^+)^{1/2} \Psi^a
\h
\Rightarrow
\h
\{\hat G^a(\us_1),\hat G^b(\us_2)\} = 0 ,
\ea
where $\Psi^a(\us)$ are fermionic ghosts which abelianize our second class
constraints as a consequence of the Poisson bracket relation
\ba\nl
\{\Psi^a(\us_1),\Psi^b(\us_2)\} = -iC^{ab}\da .
\ea

It turns out, that the constraint algebra is much more simple, if we work
not with $D^a$ and $\hat G^a$ but with $\hat T^\alpha$ given by
\ba\nl
\hat T^\alpha &=&(p^+)^{-1/2}\bigl [(\s^a v^+)^\alpha D_a +
(\not p \s^+ \s^a v^-)^\alpha \hat G_a \bigr ]
\\ \nl
&=&(p^+)^{1/2}D^\alpha + (\not p \s^+ \s^a v^-)^\alpha \Psi_a .
\ea

After the introduction of the auxiliary fermionic variables $\Psi^a$,
we have to modify some of the constraints, to preserve their first class
property. Namely $T_j$, $I^{ab}$ and $I^{-a}$ change as follows
\ba\nl
\hat T_j &=& T_j + \frac{i}{2}C^{ab}\Psi_a \p_j \Psi_b ,
\\ \nl
\hat I^{ab} &=& I^{ab} + J^{ab},
\h
J^{ab}=\int d^p \s j^{ab}(\us),
\h
j^{ab}=\frac{i}{4}(v^-\s_c\s^{ab}\s^+\s_d v^-)\Psi^c\Psi^d,
\\ \nl
\hat I^{-a}&=&I^{-a} + J^{-a},
\h
J^{-a}=\int d^p\s j^{-a}(\us),
\h
j^{-a}=-(p^+)^{-1}j^{ab}p_b .
\ea
As a consequence, we can write down the Hamiltonian for the considered model
in the form:
\ba\nl
H=\int d^p\s\bigl [\la^0 T_0(\us)+\la^j \hat T_j(\us)+
\la^\alpha\hat T_\alpha(\us)\bigr ] +
\\ \nl
\la_{ab}\hat I^{ab}+\la_{-+}I^{-+}+\la_{+a}I^{+a}+\la_{-a}\hat I^{-a} .
\ea
The constraints entering $H$ are all first class, irreducible and Lorentz-
covariant. Their algebra reads (only the non-zero Poisson brackets are
written):
\ba\nl
\{T_0(\us_1),\hat T_j(\us_2)\}&=&\bigl (T_0(\us_1)+T_0(\us_2)\bigr )\p_j \da ,
\\ \nl
\{\hat T_j(\us_1),\hat T_k(\us_2)\}&=&
\bigl (\delta_j^l \hat T_k(\us_1)+\delta_k^l \hat T_j(\us_2)\bigr )\p_l \da ,
\\ \nl
\{\hat T_j(\us_1),\hat T_\alpha(\us_2)\}&=&
\bigl (\hat T_\alpha(\us_1)+\frac{1}{2}\hat T_\alpha(\us_2)\bigr )\p_j \da ,
\\ \nl
\{\hat T_\alpha(\us_1),\hat T_\beta(\us_2)\}&=&
i\s^+_{\alpha\beta}T_0 \da ,
\\ \nl
\{I^{-+},\hat T_\alpha \}&=&\frac{1}{2}\hat T_\alpha ,
\h
\{\hat I^{-a},\hat T_\alpha \}=(2p^+)^{-1}\bigl [
p^a\hat T_\alpha + (\s^+ \s^{ab}v^-)_\alpha\Psi_b T_0 \bigr ] ,
\\ \nl
\{\hat I^{ab},\hat I^{cd}\}&=&C^{bc}\hat I^{ad}-C^{ac}\hat I^{bd}+
C^{ad}\hat I^{bc}-C^{bd}\hat I^{ac} ,
\\ \nl
\{I^{-+},I^{+a}\}&=&I^{+a} ,
\h
\{I^{-+},\hat I^{-a}\}=-\hat I^{-a} ,
\\ \nl
\{\hat I^{ab},I^{+c}\}&=&C^{bc}I^{+a}-C^{ac}I^{+b} ,
\h
\{\hat I^{ab},\hat I^{-c}\}=C^{bc}\hat I^{-a}-C^{ac}\hat I^{-b} ,
\\ \nl
\{I^{+a},\hat I^{-b}\}&=&C^{ab}I^{-+} + \hat I^{ab} ,
\\ \nl
\{\hat I^{-a},\hat I^{-b}\}&=&-\int d^p\s(p^+)^{-2}j^{ab}T_0 .
\ea

Having in mind the above algebra, one can construct the 
corresponding BRST charge $\Omega$ \cite{FF} 
\ba
\label{O} 
\Omega = \Omega^{min}+\pi_M \bar {\cal P}^M , 
\h 
\{\Omega,\Omega\} = 0 , 
\h 
\Omega^* = \Omega , 
\ea 
where $M=0,j,\alpha,ab,-+,+a,-a$. 
$\Omega^{min}$ in (\ref{O}) can be written as 
\ba\nl 
\Omega^{min}&=&\Omega^{brane}+\Omega^{aux} ,
\\ \nl
\Omega^{brane}&=&\int d^p\s\{T_0\eta^0+\hat T_j\eta^j+\hat T_\alpha
\eta^\alpha + {\cal P}_0 [(\p_j\eta^j)\eta^0 + (\p_j\eta^0)\eta^j ] +
\\ \nl
&+&{\cal P}_k(\p_j\eta^k)\eta^j +
{\cal P}_\alpha [\eta^j\p_j\eta^\alpha -
\frac{1}{2}\eta^\alpha\p_j\eta^j ] -
\frac{i}{2}{\cal P}_0\eta^\alpha\s^+_{\alpha\beta}\eta^\beta\} ,
\\ \nl
\Omega^{aux}&=&
\hat I^{ab}\eta_{ab}+I^{-+}\eta_{-+}+I^{+a}\eta_{+a}+\hat I^{-a}\eta_{-a} +
\\ \nl
&+&({\cal P}^{ac}\eta^{b.}_{.c}-{\cal P}^{bc}\eta^{a.}_{.c} +
2{\cal P}^{+a}\eta^b_+ + 2{\cal P}^{-a}\eta^b_-)\eta_{ab} +
\\ \nl
&+&({\cal P}^{+a}\eta_{+a}-{\cal P}^{-a}\eta_{-a})\eta_{-+} +
({\cal P}^{-+}\eta^a_- + {\cal P}^{ab}\eta_{-b})\eta_{+a} +
\\ \nl
&+&\frac{1}{2}\int d^p\s\{{\cal P}_\alpha\eta^\alpha\eta_{-+} +
(p^+)^{-1}[p^a{\cal P}_\alpha -(\s^+\s^{ab}v^-)_\alpha\Psi_b {\cal P}_0]
\eta^\alpha\eta_{-a}-
\\ \nl
&-&(p^+)^{-2}j^{ab}{\cal P}_0\eta_{-b}\eta_{-a}\} .
\ea
These expressions for $\Omega^{brane}$ and $\Omega^{aux}$ show that we have
found a set of constraints which ensure the first rank property of the model.

$\Omega^{min}$ can be represented also in the form
\ba\nl
\Omega^{min}=\int d^p\s\bigl [\bigl (T_0+\frac{1}{2}T_0^{gh}\bigr )\eta^0
+\bigl (\hat T_j+\frac{1}{2}T_j^{gh}\bigr )\eta^j
+\bigl (\hat T_\alpha +\frac{1}{2}T_\alpha^{gh}\bigr )\eta^\alpha\bigr ]+
\\ \nl
+\bigl (\hat I^{ab}+\frac{1}{2}I^{ab}_{gh}\bigr )\eta_{ab}
+\bigl (I^{-+}+\frac{1}{2}I^{-+}_{gh}\bigr )\eta_{-+}
+\bigl (I^{+a}+\frac{1}{2}I^{+a}_{gh}\bigr )\eta_{+a}
+\bigl (\hat I^{-a}+\frac{1}{2}I^{-a}_{gh}\bigr )\eta_{-a}+
\\ \nl
+\int d^p\s\p_j\Bigl (\frac{1}{2}{\cal P}_k\eta^k\eta^j
+\frac{1}{4}{\cal P}_\alpha\eta^\alpha\eta^j\Bigr ) .
\ea
Here a super(sub)script $gh$ is used for the ghost part of the total gauge
generators
\ba\nl
\nl
G^{tot}=\{\Omega,{\cal P}\}=\{\Omega^{min},{\cal P}\}=G+G^{gh} .
\ea
We recall that the Poisson bracket algebras of $G^{tot}$ and $G$ coincide
for first rank systems. The manifest expressions for $G^{gh}$ are:
\ba\nl
T_0^{gh}&=&2{\cal P}_0\p_j\eta^j+\bigl (\p_j{\cal P}\bigr )\eta^j ,
\\ \nl
T_j^{gh}&=&2{\cal P}_0\p_j\eta^0+\bigl (\p_j{\cal P}_0\bigr )\eta^0+
{\cal P}_j\p_k\eta^k+{\cal P}_k\p_j\eta^k+\bigl (\p_k{\cal P}_j\bigr )\eta^k
\\ \nl
&+&\frac{3}{2}{\cal P}_\alpha\p_j\eta^\alpha
+\frac{1}{2}\bigl (\p_j{\cal P}_\alpha\bigr )\eta^\alpha ,
\\ \nl
T_\alpha^{gh}&=&-\frac{3}{2}{\cal P}_\alpha\p_j\eta^j
-\bigl (\p_j{\cal P}_\alpha\bigr )\eta^j
-i{\cal P}_0\s^+_{\alpha\beta}\eta^\beta +
\\ \nl
&+&\frac{1}{2}{\cal P}_\alpha\eta_{-+}+(2p^+)^{-1}\Bigl [p^a{\cal P}_\alpha
-(\s^+\s^{ab}v^-)_\alpha\Psi_b{\cal P}_0\Bigr ]\eta_{-a} ,
\\ \nl
I^{ab}_{gh}&=&2\bigl ({\cal P}^{ac}\eta^{b.}_{.c}-{\cal P}^{bc}\eta^{a.}_{.c}
\bigr )+\bigl ({\cal P}^{+a}\eta^b_+-{\cal P}^{+b}\eta^a_+\bigr )+
\bigl ({\cal P}^{-a}\eta^b_--{\cal P}^{-b}\eta^a_-\bigr ) ,
\\ \nl
I^{-+}_{gh}&=&{\cal P}^{+a}\eta_{+a}-{\cal P}^{-a}\eta_{-a}+
\frac{1}{2}\int d^p\s{\cal P}_\alpha\eta^\alpha ,
\\ \nl
I^{+a}_{gh}&=&2{\cal P}^{+b}\eta^{a.}_{.b}-{\cal P}^{+a}\eta_{-+}+
{\cal P}^{-+}\eta^a_-+{\cal P}^{ab}\eta_{-b} ,
\\ \nl
I^{-a}_{gh}&=&2{\cal P}^{-b}\eta^{a.}_{.b}+{\cal P}^{-a}\eta_{-+}-
{\cal P}^{-+}\eta^a_++{\cal P}^{ab}\eta_{+b}+
\\ \nl
&+&\int d^p\s\Bigl \{(2p^+)^{-1}\Bigl [p^a{\cal P}_\alpha-
(\s^+\s^{ab}v^-)_\alpha\Psi_b{\cal P}_0\Bigr ]\eta^\alpha-
(p^+)^{-2}j^{ab}{\cal P}_0 \eta_{-b}\Bigr \} .
\ea
Up to now, we introduced canonicaly conjugated ghosts
$\bigl (\eta^M,{\cal P}_M\bigr )$, $\bigl (\bar \eta_M,\bar {\cal P}^M\bigr)$
and momenta $\pi_M$ for the Lagrange multipliers $\lambda^M$ in the
Hamiltonian. They have Poisson brackets and Grassmann parity as follows
($\epsilon_M$ is the Grassmann parity of the corresponding 
constraint):  
\ba\nl 
\bigl \{\eta^M,{\cal P}_N\bigr \}&=&\delta^M_N , 
\h 
\epsilon (\eta^M)=\epsilon ({\cal P}_M)=\epsilon_M + 1 , \\ 
\nl 
\bigl \{\bar \eta_M,\bar {\cal P}^N \bigr \}&=&-(-1)^{\epsilon_M\epsilon_N} 
\delta^N_M , 
\h 
\epsilon (\bar\eta_M)=\epsilon (\bar {\cal P}^M)=\epsilon_M + 1 , 
\\ \nl 
\bigl \{\lambda^M,\pi_N\bigr \}&=&\delta^M_N ,
\h
\epsilon (\lambda^M)=\epsilon (\pi_M)=\epsilon_M .
\ea

The BRST-invariant Hamiltonian is
\ba\label{H}
H_{\tilde \chi}=H^{min}+\bigl \{\tilde \chi,\Omega\bigr \}=
\bigl \{\tilde \chi,\Omega\bigr \} ,
\ea
because from $H_{canonical}=0$ it follows $H^{min}=0$. In this
formula $\tilde \chi$ stands for the gauge fixing fermion
$(\tilde \chi^* = -\tilde \chi)$. We use the following representation
for the latter
\ba\nl
\tilde \chi=\chi^{min}+\bar\eta_M(\chi^M+\frac{1}{2}\rho_{(M)}\pi^M) ,
\h
\chi^{min}=\la^M{\cal P}_M ,
\ea
where $\rho_{(M)}$ are scalar parameters and we have separated the
$\pi^M$-dependence from $\chi^M$. If we adopt that $\chi^M$ does 
not depend on the ghosts $(\eta^M,{\cal P}_M)$ and 
$(\bar\eta_M,\bar {\cal P}^M)$, the Hamiltonian $H_{\tilde\chi}$ 
from (\ref{H}) takes the form 
\ba
\label{r} 
H_{\tilde\chi}&=&H_{\chi}^{min}+{\cal P}_M \bar {\cal P}^M - 
\pi_M(\chi^M+\frac{1}{2}\rho_{(M)}\pi^M)+ \\ \nl &+&\bar\eta_M\Bigl 
[\bigl \{\chi^M,G_N\bigr \}\eta^N +\frac{1}{2}(-1)^{\epsilon_N} 
{\cal P}_Q\bigl \{\chi^M,U^Q_{NP}\bigr \}\eta^P\eta^N \Bigr ],
\ea
where
\ba\nl
H_{\chi}^{min}=\bigl \{\chi^{min},\Omega^{min}\bigr \} ,
\ea
and generaly $\bigl \{\chi^M,U^Q_{NP}\bigr \}\not=0$ as far as the structure
coeficients of the constraint algebra $U^M_{NP}$ depend on the phase-space
variables.

One can use the representation (\ref{r}) for $H_{\tilde\chi}$ to obtain the
corresponding BRST invariant Lagrangian
\ba\nl
L_{\tilde\chi}=L+L_{GH}+L_{GF} .
\ea
Here $L_{GH}$ stands for the ghost part and $L_{GF}$ - for the gauge fixing
part of the Lagrangian. If one does not intend to pass to the Lagrangian
formalism, one may restrict oneself to
the minimal sector $\bigl (\Omega^{min},\chi^{min},H_\chi^{min}\bigr )$.
In particular, this means that Lagrange multipliers are not considered as
dynamical variables anymore. 
With this particular gauge choice, $H_\chi^{min}$
is a linear combination of the total constraints
\ba\nl
H_\chi^{min}&=&H_{brane}^{min}+H_{aux}^{min}=
\\ \nl
&=&\int d^p\s\Bigl [\La^0 T_0^{tot}(\us)+\La^j T_j^{tot}(\us)+
\La^\alpha T_\alpha^{tot}(\us)\Bigr ] +
\\ \nl
&+&\La_{ab}I^{ab}_{tot}+\La_{-+}I^{-+}_{tot}+\La_{+a}I^{+a}_{tot}+
\La_{-a}I^{-a}_{tot} ,
\ea
and we can treat here the Lagrange multipliers $\La^0,...,\La_{-a}$
as constants. Of course, this does not fix the gauge completely.
\section{\bf Comments and conclusions}
\hspace{1cm}
The introduced harmonic variables $u_\mu^a$, $v_\alpha^\pm$ helped
us to construct $SO(1,1)\times SO(8)$ covariant quantities
($\s^a ,p^+ $, etc.) from $SO(1,9)$-covariant ones. We note here
that this is an invertible operation. For any Lorentz vector 
$A_\mu$ we have 
\ba\nl 
A_\nu = - u_\nu^+ A^- - u_\nu^- A^+ + 
u_\nu^a A_a , 
\ea 
where 
\ba\nl 
A^a=u_\nu^a A^\nu 
\h ,
\h 
A^\pm=u_\nu^\pm A^\nu .
\ea
For Lorentz spinors the reversibility is demonstrated
in the equality (\ref{rev}) for example.

To ensure that the harmonics and their conjugate momenta are pure
gauge degrees of freedom, we have to consider as physical
observables only such functions on the phase space which do not
carry any $SO(1,1)\times SO(8)$ indices.  More precisely, these
functions are defined by the following expansion
\ba\nl
F(y,u,v;p_y,p_u,p_v)=\sum_{}\bigl [u_{\nu_1}^{a_1}...u_{\nu_k}^{a_k}
p_{u\nu_{k+1}}^{a_{k+1}}...p_{u\nu_{k+l}}^{a_{k+l}}\bigr ]_{SO(8)
singlet}\\ \nl
v_{\alpha_1}^+...v_{\alpha_m}^+v_{\alpha_{m+1}}^-...v_{\alpha_{m+n}}^-
p_v^{+\beta_1}...p_v^{+\beta_r}p_v^{-\beta_{r+1}}...p_v^{-\beta_{m-n+r}}\\
\nl
F_{\beta_1...\beta_{m-n+r}}^{\alpha_1...\alpha_{m+n},\nu_1...\nu_{k+l}}
(y,p_y) ,
\ea
where $(y,p_y)$ are the non-harmonic phase space conjugated pairs.

In this letter we begin the investigation of a $p$-brane model with
$N=1$ supersymmetry in 10-dimensional flat space-time. Starting
with a Hamiltonian which is a linear combination of first and mixed
(first and second) class constraints, we succeed to obtain a new 
one, which is a linear combination of first class, irreducible and 
Lorentz-covariant constraints only. This is done with the help of
the introduced auxiliary harmonic variables. Then we give manifest
expressions for the classical BRST charge, the corresponding total
constraints and BRST-invariant Hamiltonian. It turns out, that in
the given formulation our model is a first rank dynamical system.
The problems of Lagrangian formulation, finding solutions of the
classical equations of motion and quantization will be considered
in the second part of the paper.

\vspace*{1cm}
{\bf Acknowledgments}

The author would like to thank A. Pashnev
and B. Dimitrov for critical reading of the manuscript.

\vspace*{1cm}
{\Large{\bf Appendix}}
\vspace*{.5cm}
\hspace{1cm}

We briefly describe here our 10-dimensional conventions. Dirac $\gamma$-
matrices obey
\ba\nl
\Gamma_\mu\Gamma_\nu+\Gamma_\nu\Gamma_\mu=2\eta_{\mu\nu}
\ea
and are taken in the representation
\ba\nl
\Gamma^\mu = \left(\begin{array}{cc}0&(\s^\mu)_\alpha^{\dot\beta}\\
(\tilde\s^\mu)^\beta_{\dot\alpha}&0\end{array}\right)
\h .
\ea
$\Gamma^{11}$ and charge conjugation matrix $C_{10}$ are given by
\ba\nl
\Gamma^{11}=\Gamma^0\Gamma^1 ... \Gamma^9 =
\left(\begin{array}{cc}\delta_\alpha^\beta &0\\
0&-\delta_{\dot\alpha}^{\dot\beta}\end{array}\right)
\h ,
\ea
\ba \nl
C_{10}=\left(\begin{array}{cc}0&C^{\alpha\dot\beta}\\
(-C)^{\dot\alpha\beta}&0\end{array}\right)
\h ,
\ea
and the indices of right and left Majorana-Weyl fermions are raised as
\ba\nl
\psi^\alpha=C^{\alpha\dot\beta}\psi_{\dot\beta}
\h ,\h
\phi^{\dot\alpha}=(-C)^{\dot\alpha\beta}\phi_\beta .
\ea

We use $D=10$ $\s$-matrices with undotted indices
\ba\nl
(\s^\mu)^{\alpha\beta}=C^{\alpha\dot\alpha}(\tilde\s^\mu)_{\dot\alpha}^\beta
\h ,\h
(\s^\mu)_{\alpha\beta}=(-C)^{-1}_{\beta\dot\beta}(\s^\mu)_\alpha^{\dot\beta}
\h ,
\ea
and the notation
\ba\nl
\s^{\mu_1...\mu_n}\equiv\s^{[\mu_1}...\s^{\mu_n]}
\ea
for their antisymmetrized products.

From the corresponding properties of $D=10$ $\gamma$-matrices, it follows:
\ba\nl
(\s^\mu)_{\alpha\gamma}(\s^\nu)^{\gamma\beta}+
(\s^\nu)_{\alpha\gamma}(\s^\mu)^{\gamma\beta}=
-2\delta_\alpha^\beta\eta^{\mu\nu}\h ,
\\ \nl
(\s_{\mu_1 ... \mu_{2s+1}})^{\alpha\beta}=
(-1)^s(\s_{\mu_1 ... \mu_{2s+1}})^{\beta\alpha}\h ,
\\ \nl
\s^\mu\s^{\nu_1 ... \nu_n}=\s^{\mu\nu_1 ... \nu_n}+
\sum_{k=1}^{n}(-1)^k\eta^{\mu\nu_k}
\s^{\nu_1 ... \nu_{k-1}\nu_{k+1} ... \nu_n}.
\ea

The Fierz identity for the $\s$-matrices reads:
\ba\nl
(\s_\mu)^{\alpha\beta}(\s^\mu)^{\gamma\delta}+
(\s_\mu)^{\beta\gamma}(\s^\mu)^{\alpha\delta}+
(\s_\mu)^{\gamma\alpha}(\s^\mu)^{\beta\delta} = 0 .
\ea

\vspace*{.5cm}


\end{document}